\documentclass{cimento}



\usepackage{graphicx}  

\title{Isoscalar Giant Monopole Resonance in Mo-100:  TDHF calculations and ground state deformation}
\author{P.~D.~Stevenson\from{ins:x}\ETC,
Abhishek\from{ins:x}
        \atque
Y.~Shi\from{ins:y}}
\instlist{
  \inst{ins:x} School of Mathematics and Physics, University of Surrey, Guildford, Surrey, GU2 7XH, UK
  \inst{ins:y} Department of Physics, Harbin Institute of Technology, Harbin 150001, China}

\begin{document}

\maketitle

\begin{abstract}
The isoscalar giant monopole resonance in molybdenum-100 has been measured in recent alpha scattering experiments and its strength function extracted.  By performing time-dependent Hartree-Fock calculations with Skyrme effective interactions, we are able to reproduce the gross features of the observed strength.  The details of the structure are found to be dependent on the ground state shape.  We use a measure of fit to determine which ground state shape correlates with the observed strength and find evidence for triaxiality in this nucleus.
\end{abstract}

\section{Introduction}
Recent experimental work at RCNP Osaka probed the isoscalar monopole (ISM) strength in open-shell isotopes near A=90 \cite{How20}, showing strength functions suggestive of the onset of deformation with increasing netron number above the N=50 magic number.  A microscopic analysis based on RPA built upon axial ground state shape showed how deformation leads to a characteristic two-peaked structure, suggesting contribution from the coupling between monopole and quadrupole vibrations \cite{Col20}.

We performed time-dependent shape-constrained Hartree-Fock calculations of the ISM resonance in molybdenum-100 to further explore the link between the structure of this nucleus and its ISM strength function \cite{Shi23}.  We found a clear variation between predicted strength functions and underlying structure, as well as an explicit coupling between the monopole and quadrupole excitations.  In that work, we concluded that the shape of the Mo-100 ground state was likely triaxial with shape parameters around $\beta_2=0.28$, $\gamma=30^\circ$.  This conclusion was based on a visual comparison between calculated strength functions and experimental data.

In the present contribution, we make a more quantitative interpretation of our calculations, and reinforce our previous conclusions as a result.

\section{Method}

The calculation of giant resonances via the time-dependent Hartree-Fock (TDHF) method is well-descripbed in the literature and we refer readers to previous publications for details \cite{Shi23,sky3d,ijmpe,maruhn,burrello,simenel}.  In the calculations presented here and in \cite{Shi23}, a new triaxial TDHF code with shape constraints is used \cite{hit3d} to choose various shape-constrained ground states upon which to excite an isoscalar monopole excitation.  The strength function is obtained, as usual in this approach, from the Fourier transform of the time-dependent expectation value of the ISM operator.

For each calculated strength function at each chosen ground state deformation, a comparison to experimental data (from \cite{How20,Howardarxiv}) was made with the following weighted error

\begin{equation}
  W = \sqrt{\frac{1}{N}\frac{\sum_{i=1}^N \Delta_i^{-1}(S_{i,e}-S_{i,t})^2}{\sum_{i=1}^N\Delta_i^{-1}}}
  \end{equation}
where $S_{i,e}$ is the $i^{\mathrm th}$ experimental data point for the strength function, $S_{i,t}$ is the corresponding theoretical value at the same energy and $\Delta_i$ is the experimental error at that point.  $N$ is the total number of experimental data points.  Theoretical values are obtained at the experimentally measured energies by interpolation from the equally-spaced data points coming from the Fourier transform.

\section{Results and Discussion}

As an example calculation, we took the results from the previous work \cite{Shi23} using the SLy4 Skyrme effective interaction \cite{sly4}, where we had performed calculations at the following sets of deformations: $(\beta_2,\gamma)=\{(0.17,0^\circ),(0.2,0^\circ),(0.2,30^\circ),(0.28,30^\circ)\}$.

\begin{figure}[tbh]
\includegraphics[width=\textwidth]{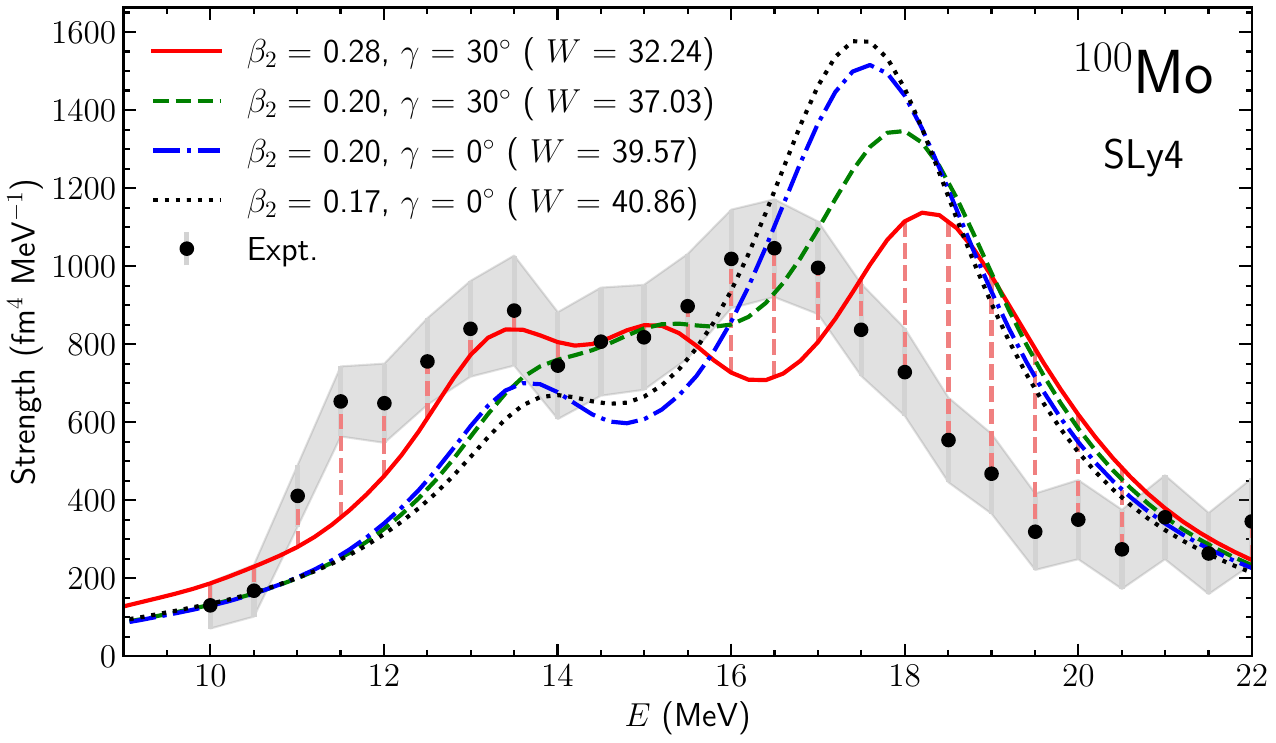}
\caption{Theoretical (lines) and experimental (points with errorbars and shaded error envelope) strength functions for the isoscalar giant monopole resonance in Mo-100.  Theoretical values are labelled with ground state deformation paramters $\beta_2$ and $\gamma$ along with weighted error values $W$.}\label{fig}
\end{figure}

Figure \ref{fig} shows the results of the calculations along with the experimental data, with errors, corresponding to Fig. 5 of \cite{Shi23}.  Also shown for the case of $(\beta_2,\gamma)=(0.28,30^\circ)$ are dashed vertical lines showing the difference between each experimental point and the theoretical calculation.  These are the differences $S_{i,e}-S_{i,t}$ that contribute to the weighted error, $W$.  The figure legend shows the different deformation parameters used in the ground state shape constraint and, in parentheses, the numerical value of the weighted error.

Since the lowest values of $W$ correspond to the best fits, it is seen that the the result corresponding to the ground state shape with $(\beta_2,\gamma)=(0.28,30^\circ)$ agrees best of all shapes tried.  This agrees with the visual assessment in our original paper \cite{Shi23}.

One can imagine use of an optimisation algorithm to minimise $W$ through variation of $\beta_2$ and $\gamma$, though each TDHF run is time consuming, and usually some level of human sanity-checking of output is required.  Nevertheless, use of a weighted fit provides a robust basis for further studies along these lines.

\acknowledgments
YS acknowledges the HPC Studo of th Physics Department at the Harbin Institute of Technology for computing resources allocated through INSPUR-HPC@PHY.HIT.  The University of Surrey authors acknowledge support from STFC through grant ST/V001108/1.


\begin{thebibliography}{0}
\bibitem{How20} \BY{Howard K., Garg U., Itoh M., \textit{et al.}}
  \IN{Phys. Lett. B}{807}{2020}{135608}
\bibitem{Shi23} \BY{Shi Y. \atque Stevenson P.}
\IN{Chin. Phys. C}{47}{2023}{034105}
\bibitem{Col20} \BY{Col\`o G., Gambacurta D., Kleinig W., Kvasil J., Nesterenko, V. \atque Pastore A.}
  \IN{Phys. Lett. B}{811}{2020}{135940}
\bibitem{sky3d} \BY{Maruhn J. A., Reinhard, P.-G., Stevenson, P. D., \atque Umar. S.}
  \IN{Comput. Phys. Commun.}{185}{2014}{2195}
\bibitem{ijmpe} \BY{Stevenson P. D., Strayer M. R, Rikovska Stone J., \atque Newton, W. G.}
  \IN{Int. J. Mod. Phys. E}{13}{2004}{181}
\bibitem{maruhn} \BY{Maruhn J. A., Reinhard P.-G., Stevenson P. D., Rikovska Stone J., \atque Strayer M. R.}
  \IN{Phys. Rev. C}{71}{2005}{064328}
\bibitem{burrello} \BY{Burrello S., Colonna M., Col\`o G., Lacroix D., Roca-Maza X., Scamps G., \atque Zheng H.}
  \IN{Phys. Rev. C}{99}{2019}{054314}
\bibitem{simenel}\BY{Simenel C.}
  \IN{Eur. Phys. J. A}{48}{2012}{152}
\bibitem{hit3d} \BY{Shi Y., Stevenson, P. D., \atque Hinohara N.}
  \TITLE{in preparation}
\bibitem{Howardarxiv} \BY{Howard K.}
  \TITLE{PhD Thesis, University of Notre Dame} (2020)
\bibitem{sly4} \BY{Chabanat E., Bonche P., Haensel P., Meyer J., \atque Schaeffer R.}
  \TITLE{Proceedings of International Workshop on Research with Fission Fragments}, World Scientific, Singapore (1996)



\end{thebibliography}
\end{document}